\documentclass[
submission
]{dmtcs-episciences}

\usepackage{amsmath,amssymb,amsfonts,amsthm,epsfig,color,graphicx,eucal,mathrsfs}
\usepackage{picins}
\usepackage{subfig}
\usepackage[utf8]{inputenc}
\newtheorem{definition}{Definition}
\newtheorem{remark}{Remark}
\newtheorem{theorem}{Theorem}
\usepackage[round]{natbib}

\author{Payal\affiliationmark{1}\thanks{Email: payal.dtu@gmail.com}
  \and Sangita Kansal \affiliationmark{1} \thanks{(Corresponding Author)Email: sangita\_kansal15@rediffmail.com}
  }
\title{Domination in Signed Petri Net}
\affiliation{
  	Department of Applied Mathematics,Delhi Technological University
  }
\keywords{Petri net; Signed graph; Domination in graph}
\received{2020-01-14}
\revised{2020-01-14}
\accepted{2020-02-14}
\begin{document}
\publicationdetails{VOL}{2020}{ISS}{NUM}{SUBM}
\maketitle
\begin{abstract}
 In this paper, domination in Signed Petri net(SPN) has been introduced.We identify some of the Petri net structures where a dominating set can exist.Applications of producer consumer problem, searching of food by bees and finding similarity in research papers are given to understand the areas where the proposed theory can be used. 
\end{abstract}

\section{Introduction}

The foundations of theory of domination can be traced back to the chess problem of finding the minimum number of queens required such that all the squares are either occupied or can be attacked by a queen\cite{CF}.The applications of theory of domination includes communication network problems, facility location problem, routings, etc.\cite{Sas,Pre} .The domination in graphs and signed graphs have been well studied by various authors in different forms viz. Roman domination, double domination, total domination, signed domination, signed total domination etc.\cite{Har1,Hay,BD,As,Pal,Bli,Sam,Boh}.
\par In \cite{Pay}, the authors introduced the concept of Signed Petri net(SPN) by utilizing the properties of signed graph and Petri net.SPNs are capable of modeling a large variety of systems and is prefered over a Petri net due to the presence of two types of tokens in it, positive and negative, which are distinguishable. Other advantages of using SPN over previously defined extensions of PN are the ability to assign sign to vertices of an SPN which is further utilized to introduce the concept of a balanced SPN as defined in \cite{Pay}.Further, in comparison to a signed graph, SPN is advantageous since a single SPN can be used to represent various signed graphs by simply varying the marking of SPN due to firing of a sequence of transitions.Thus, we need to analyse one SPN in order to infer about all possible signed graph structures that can be formed for a fixed number of vertices.
\par
As SPN is a bipartite graph, it can be used to develop the theory of domination for dynamic systems as such a theory is not prevalent for Petri nets.
\section{Basic Definitions}

\subsection{Petri Net (PN)}

A \textit{Petri net }\cite{Jen} is a 5-tuple $ N=(P,T,I^{-},I^{+},\mu_0)$,where
\begin{enumerate}
	\item 	P is the finite, non-empty set of places.
	\item 	T is the finite, non-empty set of transitions.
	\item 	$ P\cap T =\emptyset$.
	\item 	$ I^{-},I^{+} :(P\times T)\rightarrow \mathbb{N}$ where $\mathbb{N}$ is the set of non-negative integers, are called negative and positive incidence functions respectively.
	\item 	$\forall p \in P,\exists \ t \in T $ such that $I^{-}(p,t) \neq 0 \ or \ I^{+}(p,t) \neq 0$,and\\
	$\forall t \in T,\exists \ p \in P $ such that $I^{-}(p,t) \neq 0 \ or \ I^{+}(p,t) \neq 0$
	
	\item 	$\mu_0 :P\rightarrow \mathbb{N} $ is the initial marking which gives the initial distribution of tokens in places.
	
	The arc set of the Petri net $N$ is defined as:
	$$E=\{(p,t):I^{-}(p,t)>0\} \cup \{(t,p):I^{+}(p,t)>0\}$$
	
\end{enumerate}

\subsection{Signed Petri Net }
\label{S_1}

\begin{definition}{\textbf{Signed Petri Net (SPN)}}
	
	A Signed Petri Net \cite{Pay} is defined as a 3-tuple $N^{*}=(N',\sigma,\mu_0)$ ,where 
	\begin{enumerate}
		\item $N'=(P,T,I^{-},I^{+})$ is a Petri net structure. 
		\item $\sigma :E \rightarrow \{+,-\}$, where $E$ is the arc set of $N'$.An arc is called a positive or negative arc respectively according to the sign $+$ or  $-$ assigned to it using the function $\sigma$.
		\item  $\mu_0=(\mu^{+}_0,\mu^{-}_0)$ is the initial marking of SPN where
		\begin{enumerate}
			\item $\mu_0^{+}:P\rightarrow \mathbb{N}$ gives the initial distribution of positive tokens in the places,called positive marking of SPN.
			\item $\mu_0^{-}:P\rightarrow \mathbb{N}$  gives the initial distribution of negative tokens in the places,called negative marking of SPN.
		\end{enumerate}
	\end{enumerate}
\end{definition}

Thus,a marking in SPN can be represented as a vector $\mu=(\mu^{+},\mu^{-})$ with $\mu^+,\mu^- \in \mathbb{N}^n,n=|P|$ such that $\mu(p_i)=(\mu^{+}(p_i),\mu^{-}(p_i)) \ \forall \ p_i \in P$.\\
\par Graphically, positive and negative arcs in an SPN are represented by solid and dotted lines respectively.A positive token is represented by a filled circle and a negative token by an open circle.\\
An SPN is said to be \textit{negative} if all of its arcs are negative in sign.

\begin{remark}
	$N''=(N',\sigma)$ is called an SPN structure where $N'$ is a PN structure and $\sigma :E \rightarrow \{+,-\}$.
\end{remark}

\subsubsection{Execution Rules for Signed Petri Net}
\label{prop}
Similar to a Petri net, the execution of an SPN depends on the distribution of tokens in its places.The execution takes place by firing of a transition.A transition may fire if it is enabled.\\
A transition $t$ in an SPN $N^{*}$ is \textit{enabled} at a marking $\mu=(\mu^+,\mu^-)$ if \\
$$
I^{-}(p,t) \leq \mu^{+}(p) \ \forall p \in {}^{\bullet}t \ \textnormal{for which} \ \sigma(p,t) = + $$
$$	 I^{-}(p,t) \leq \mu^{-}(p) \ \forall p \in {}^{\bullet}t \ \textnormal{for which} \ \sigma(p,t) = - 
$$
An enabled transition $t$ may \textit{fire} at $\mu=(\mu^+,\mu^-)$ provided $\exists p_k \in t^{\bullet}$ such that:
$$\sigma(t,p_k)=
\begin{cases}
+ & \text{if}\  \sigma(p,t)=+ \ \forall p \in {}^{\bullet}t \\
- & \text{if}\  \sigma(p,t)=- \ \forall p \in {}^{\bullet}t \\
+ \ \text{or} \ - & \text{if}\  \sigma(p,t)=+ \ \text{for some} \ p \in {}^{\bullet}t \ and \ - \ \text{for some} \ p \in {}^{\bullet}t \\
\end{cases}
$$
After firing,it yields a new marking $\mu_1=(\mu_1^+,\mu_1^-)$ given by the rule:
$$\mu_1^+(p)=\mu^+(p) - I^{-}(p,t) +  I^{+}(p,t) \ \forall p \in P \quad \textnormal{where (p,t) \& (t,p) are positive arcs,}$$ $$ \textnormal{\qquad \qquad \qquad if exist} $$
$$\mu_1^-(p)=\mu^-(p) - I^{-}(p,t) +  I^{+}(p,t) \ \forall p \in P \quad \textnormal{where (p,t) \& (t,p) are negative arcs,}$$ $$ \textnormal{\qquad \qquad \qquad if exist}$$
We say that $\mu_1$ is reachable from $\mu$ and write $\mu \stackrel{t}\to \mu_1$.We restrict the movement of positive(negative) tokens to positive(negative) arcs only.

\par
\begin{figure}[ht]
	\centering
	\subfloat[SPN with $\mu_0=((1,0,1,0),(1,0,0,0))$ ]{{\includegraphics[scale=0.35]{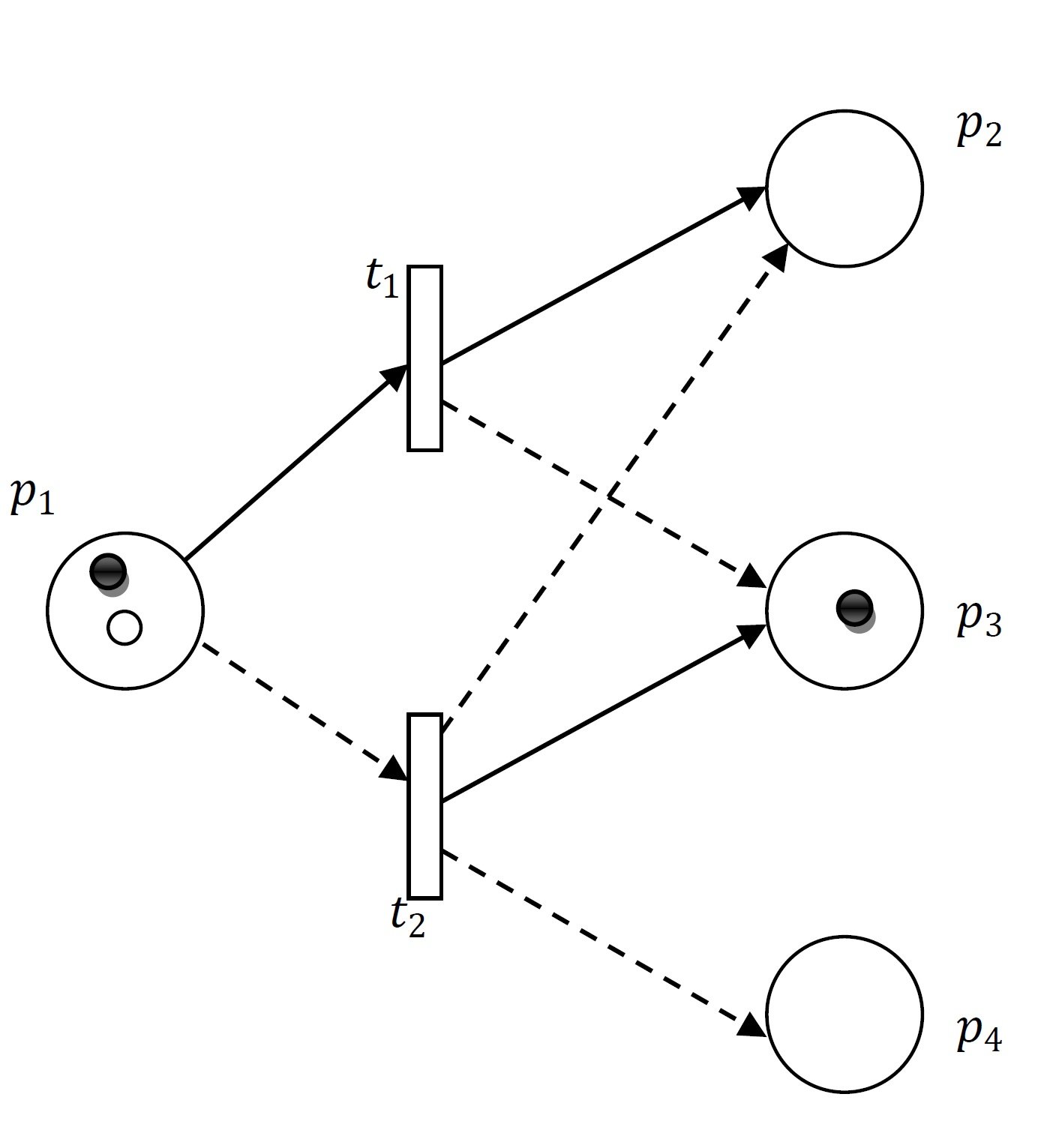} }}
	\qquad \qquad \qquad
	\subfloat[SPN with $\mu_0=((1,0,0,0),(0,0,0,0))$]{{\includegraphics[scale=0.3]{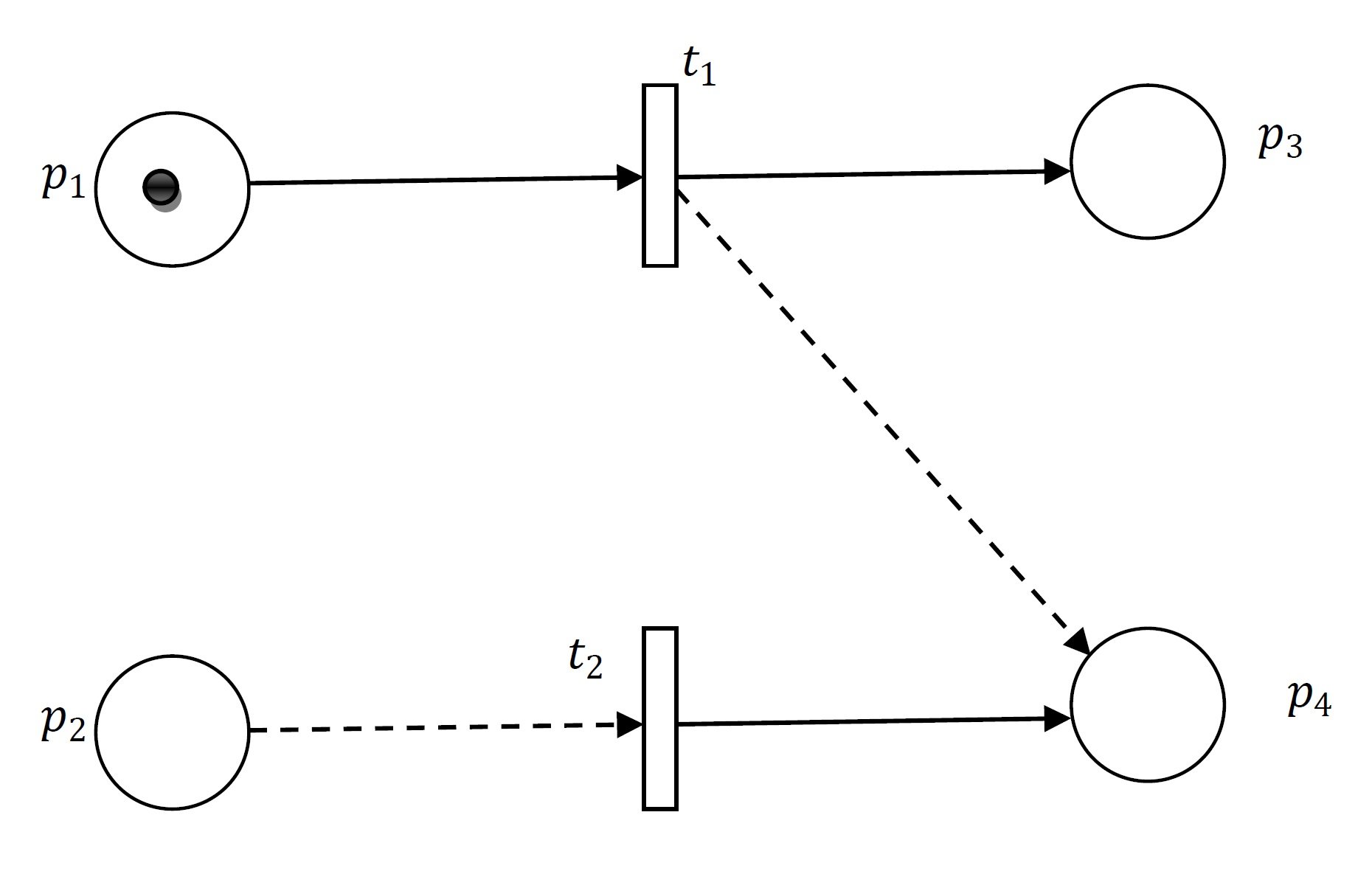} }}
	\caption{ {Execution of an SPN}}
	\label{SPN13}
\end{figure}

A marking $\mu$ is reachable from $\mu_0$ if there exists a firing sequence $\eta$ that transforms $\mu_0$ to $\mu$ and is written $\mu_0 \stackrel{\eta}\to \mu$. A \textit{firing or occurence sequence} is a sequence of transitions $\eta=t_1t_2\ldots t_k$ such that 
$$\mu_0 \stackrel{t_1}\to \mu_1\stackrel{t_2}\to \mu_2 \stackrel{t_3} \to \mu_3 \ldots \stackrel{t_k}\to \mu$$
Note that a transition $t_j,1 \leq j \leq k$ can occur more than once in the firing sequence $\eta$.\\
Let us look at the execution of an SPN with the help of an example.\\
In figure \ref{SPN13}$(a)$, \ $t_1$ and $t_2$ both are enabled at $\mu_0$. Firing of $t_1$ yields a new marking $\mu=((0,1,1,0),$$(1,0,1,0))$ and firing of $t_2$ yields a new marking $\mu=((1,0,2,0),(0,1,0,1))$.
In figure \ref{SPN13}$(b)$, \ $t_1$ is enabled,while $t_2$ is not. $t_1 $ can fire to give a new marking
$\mu=((0,0,1,0),(0,0,0,1))$.\\

\begin{definition}{\textbf{Reachability Set of Signed Petri net}}
	
	The Reachability Set $R(N^{*},\mu_0)$ of an SPN $N^{*}$ is the set of all markings of $N^*$ reachable from the initial marking $\mu_0$.
\end{definition}

\subsection{Assignment of sign to vertices of an SPN}
The vertices in an SPN can also be assigned sign.Transitions are assigned sign by product of sign of arcs (incoming and outgoing) incident on it.In Figure \ref{SPN13}(a) and \ref{SPN13}(b), all transitions are negative in sign.

Places can be assigned sign in one of the two ways:
\begin{enumerate}
	\item \textbf{With respect to arcs --}\ Sign is assigned to a place by taking product of incident arcs (incoming and outgoing) on that place.In  Figure \ref{SPN13}(a),all the places are negative in sign while in Figure \ref{SPN13}(b) $p_1$ and $p_3$ are positive in sign while $p_2$ and $p_4$ are negative in sign.
	\item \textbf{With respect to marking --}\  Sign is assigned to a place by taking product of sign of tokens in that place in the given marking.A place without token is considered to be positive.In Figure \ref{SPN13}(a), places $p_2,p_3,p_4$ are positive in sign while $p_1$ is negatively signed w.r.t. the marking $\mu_0$.In Figure \ref{SPN13}(b), all the places are positively signed w.r.t. the marking $\mu_0$.
\end{enumerate}

\textit{Remark}-- Assigning sign to places with respect to arcs doesn't utilize the most important characteristic(dynamic behaviour) of PN which is a marking.Hence, assigning sign to places with respect to marking has been used throughout the paper.
\par
\begin{figure}[h]
	\centering
	\includegraphics[scale=0.28]{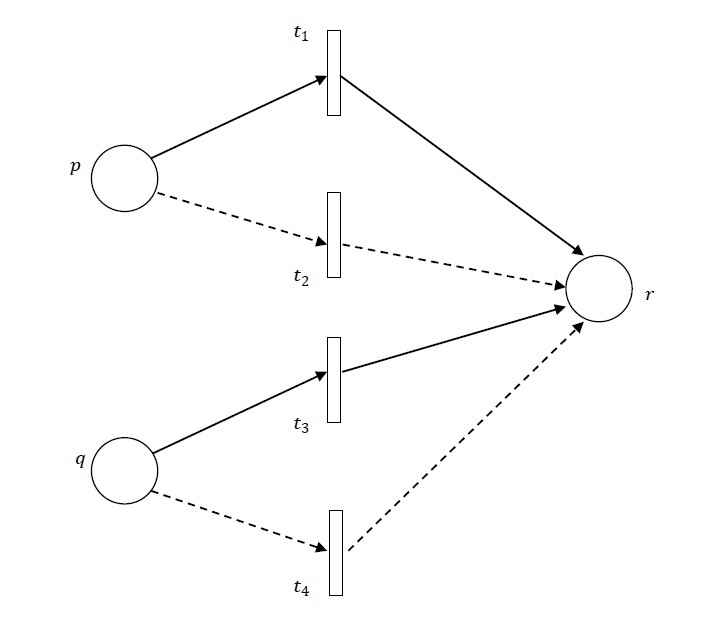}
	\caption{\textit{An SPN model to check for approval or disapproval of the product}}
	\label{fig1}
\end{figure}

An example is given which uses the concept of place sign to determine whether a product is approved or disapproved by a company.A company has to make a decision on a certain product by voting of two board members.This situation is represented in figure \ref{fig1} by modeling it with an SPN.

In the figure, $p$ and $q$ represent the board members .The transitions $t_1$ and $t_3$ will fire if positive tokens exist in places $p$ and $q$ while $t_2$ and $t_4$ transitions fire if $p$ and $q$ have negative tokens.A positive token is generated in places representing board members($p$ and $q$) if they approve the product and else if they dispprove of it , a negative token is generated.A decision on the product is reached if either both $p$ and $q$ disapproves or both approves the product.On the other hand,if one member approves the product while other rejects it,no decision is made.We can determine the decision made by the company on the basis of the sign of the place $r$ in the final marking(when place $r$ gets a token).This is shown in table \ref{table1}.

\begin{table*}[htb]
	\centering
	\caption{Decision Table for the Product} \label{table1}
	\begin{tabular}{| c |c |c |} 
		\hline
		Transition firings & Sign of place $r$ & Decision made \\
		\hline
		\hline
		$t_1 ,t_3$  &  +  &  Yes\\
		\hline
		$t_1 ,t_4$  &  -  &  No \\
		\hline
		$t_2 ,t_3$  &  -  &   No \\
		\hline
		$t_2 ,t_4$  &  +  &  Yes \\
		\hline

	\end{tabular}

\end{table*}

Thus,based on the sign of the place $r$,we can infer the following:
\begin{enumerate}
	\item The decision made about the product(The company reaches a final decison if the sign of place $r$ is positive in marking with $\mu(r) \neq 0$,else no decision is made regarding product).
	\item Whether $p$ and $q$ have same opinions or varied ones(Both the members have same opinion regarding product if the sign of place $r$ is positive,else they have varied opinion).
\end{enumerate}
If initially the places $p$ and $q$ in figure \ref{fig1} have one positive and negative token respectively, then this situation corresponds to second row in table \ref{table1}.Hence,the product will not be approved by the company.

\section{Domination Theory}
In this section, we assume an ordinary SPN (without multiple arcs) unless stated otherwise. 
\begin{definition}{\textbf{Dominating Set}}
	
	A set $D \subseteq V=PUT$ in an SPN $N^*$ is called a Dominating set with respect to a marking $\mu_1 \in R(N^*,\mu_0)$ if either all the vertices of $V$ are in $D$ or $\forall \ v \in V\backslash D$
	$$ v^\bullet \cap D \neq \emptyset \ \textnormal {and} \ \sigma(v,u)=S(v).S(u) \ \forall \ u \in v^\bullet \cap D $$
	where $S(v),S(u)$ are sign of vertices $u,v$ with respect to marking $\mu_1$.
\end{definition}
\begin{remark}
	It may be noted that sign of a transition remains same irrespective of marking of the SPN while sign of a place may vary if the marking of SPN changes.
\end{remark}
\begin{definition}{\textbf{Dominating Set with respect to a set of markings($D_M$)}}
	
	A set $D_M\subseteq V$ in an SPN $N^*$ is called a Dominating set with respect to a set of markings $M \subseteq R(N^*,\mu_0)$ if $D_M$ is a dominating set with respect to all the markings $\mu \in  M$.(Clearly,$|M|\geq 2$)
	
\end{definition}

\begin{definition}{\textbf{Dependent(or connected) Dominating Set}}
	
	A dominating set $D_M$ with respect to a set of markings $M$ in an SPN $N^*$ is called a dependent Dominating set if the markings of $M$ are all the nodes of some subtree of the reachability tree of $N^*$.
	
\end{definition}

\begin{definition}{\textbf{Minimal Dominating Set}}
	
	A dominating set $D$ is called Minimal if no proper subset of it is a dominating set or it is a dominating set with minimum number of vertices.
	
\end{definition}
\begin{remark}
	For application  purposes we would like to find minimal dependent dominating set $D$ over a maximal set of markings $M$,i.e. We try to find a maximal set of markings,$M$ over which $D$ is a minimal dominating set.
\end{remark}
\begin{theorem}
	For an SPN structure $N''$ in which all the transitions are positively signed and each place has incident(input/output) arcs of one kind only, we can find a marking $\mu$ w.r.t. which  $V\backslash A $ is a dominating set where $A$ is a set of source vertices.
\end{theorem}
\begin{proof}
	If $A=\emptyset$, then $V\backslash A=V$ is a dominating set by definition.\\
	If $A\neq \emptyset$, then we have to find a marking $\mu$ such that $V\backslash A$ is a dominating set w.r.t. $\mu$. For any $x \in V\backslash(V \backslash A)= A$,  $x^{\bullet} \cap (V\backslash A) \neq \emptyset$(Since, $x$ is a source vertex).
	Then, we find a marking $\mu$ such that $\forall y \in x^{\bullet} \cap (V\backslash A)$  \begin{equation}{\label{1}}
	\sigma(x,y) =S(x)S(y)
	\end{equation}
	where $S(x)$ and $S(y)$ are sign of vertices w.r.t. $\mu$.
	
	Let $y \in x^{\bullet} \cap (V\backslash A) $. Now, two cases arise:
	
	\begin{enumerate}
		\item \textbf{$x$ is a place.}\\
		Therefore, $y$ is a transition. 
		$\implies S(y)= + $ (By hypothesis).
		\begin{enumerate}
			\item Now, if $\sigma(x,y)=+$.In this case, for equation \ref{1} to hold ;$S(x)=+$.Therefore,
			$\mu(x)=(\mu^+(x),\mu^-(x))$ where $\mu^+(x) \in \mathbb{N} \cup \{0\}\ \& \ \mu^-(x) \in 2\mathbb{N} \cup \{0\}. $
			\item However, if $\sigma(x,y)=-$.In this case, for equation \ref{1} to hold ;$S(x)=-$.Therefore,			
			$\mu(x)=(\mu^+(x),\mu^-(x))$ where $\mu^+(x) \in \mathbb{N} \cup \{0\}\  \& \ \mu^-(x) \in \mathbb{N} \backslash 2\mathbb{N}. $
		\end{enumerate}
		\item \textbf{$x$ is a transition.}\\
		Therefore, $y$ is a place.Then, $S(x)= + $ (By hypothesis).Now, $\sigma(x,y)$ can be positive or negative.
		\begin{enumerate}
			\item If $\sigma(x,y)=+$ then, for equation \ref{1} to hold ;$S(y)=+$ and hence,
			$\mu(y)=(\mu^+(y),\mu^-(y))$ where $\mu^+(y) \in \mathbb{N} \cup \{0\}\ \& \ \mu^-(y) \in 2\mathbb{N} \cup \{0\}. $
			\item However, if $\sigma(x,y)=-$  then, for equation \ref{1} to hold ;$S(y)=-$ and hence,
			$\mu(y)=(\mu^+(y),\mu^-(y))$ where $\mu^+(y) \in \mathbb{N} \cup \{0\}\ \& \ \mu^-(y) \in \mathbb{N} \backslash 2\mathbb{N}. $
		\end{enumerate}
	\end{enumerate}
	Hence, for all $\ p_i \ \in \ A \cup \{z\ \in \ V\backslash A \ | \ z \in A^{\bullet}\}\ $, $\mu^+(p_i) \ \in \ \mathbb{N} \cup \{0\} \ \& \\ $
	$\mu^-(p_i) \ \in \ 
	\begin{cases}
	2\mathbb{N} \cup \{0\} &, \  \textnormal{if} \ p_i \ \textnormal{has positive incident arcs}. \\
	\mathbb{N} \backslash 2\mathbb{N} &, \  \textnormal{if} \ p_i \ \textnormal{has negative incident arcs}. \\
	\end{cases}
	$\\
	All the remaining places can have any number of positive and negative tokens without any restrictions.
\end{proof}

\begin{theorem}
	If an SPN structure $N''$ with no source/sink vertices and in which any place has only one type of incident arcs then, we can find a marking $\mu$ such that $P$ and $T$ are dominating sets w.r.t. $\mu$, provided all the transitions are of same sign.
\end{theorem}
\begin{proof}
	Since all the transitions are of same sign,two cases arise:
	\begin{enumerate}
		\item \textbf{Transitions are positively signed.}\\
		We find a marking $\mu$ w.r.t. which $P$ and $T$ are dominating sets.
		\begin{enumerate}
			\item \textbf{$P$ is a dominating set}.\\
			Let $t \in V\backslash P  $,therefore,  $t^{\bullet} \cap P \neq \emptyset$ (since there are no source/sink vertices). We need to find a marking $\mu$ such that $\forall \ p\ \in \ t^{\bullet} \cap P$;
			\begin{equation}{\label{2}}
			\sigma(t,p)=S(t)S(p)
			\end{equation}
			
			where $S(p)$ is the sign of place $p$ w.r.t. marking $\mu$ and $S(t)$ is the sign of transition $t$.\\
			Let $p \in t^{\bullet} \cap P$.
			Then,
			\begin{enumerate}
				\item $\sigma(t,p)= + $ \\
				Now, since $S(t)=+ \ \forall t \in T$.Then, for equation \ref{2} to hold $S(p)=+$. Then, $\mu_(p)=(\mu^+(p),\mu^-(p))$ where $\mu^+(p) \in \mathbb{N} \cup \{0\},\mu^-(p) \in 2\mathbb{N} \cup \{0\}$.
				\item $\sigma(t,p)= - $ \\
				Now, since $S(t)=+ \ \forall t \in T$.Then, for equation \ref{2} to hold $S(p)=-$. Then, $\mu_(p)=(\mu^+(p),\mu^-(p))$ where $\mu^+(p) \in \mathbb{N} \cup \{0\},\mu^-(p) \in \mathbb{N} \backslash 2\mathbb{N} $.
			\end{enumerate}
			
			\item \textbf{$T$ is a dominating set}.\\
			Let $p \in V\backslash T$,therefore, $p^{\bullet} \cap T \neq \emptyset$ (since there are no source/sink vertices). We need to find a marking $\mu$ such that $\forall \ t\ \in \ p^{\bullet} \cap T$;
			\begin{equation}{\label{3}}
			\sigma(p,t)=S(p)S(t)
			\end{equation}
			
			where $S(p)$ is sign of place $p$ w.r.t. marking $\mu$ and $S(t)$ is the sign of transition $t$.\\
			Let $t \ \in p^{\bullet} \cap T$.Then,
			\begin{enumerate}
				\item $\sigma(p,t)= + $\\
				Now, since $S(t)=+ $. Then, for equation \ref{3} to hold $S(p)=+$. Then, $\mu_(p)=(\mu^+(p),\mu^-(p))$ where $\mu^+(p) \in \mathbb{N} \cup \{0\},\mu^-(p) \in 2\mathbb{N} \cup \{0\}$.
				\item $\sigma(p,t)= - $\\
				Now, since $S(t)=+ $. Then, for equation \ref{3} to hold $S(p)=-$. Then, $\mu_(p)=(\mu^+(p),\mu^-(p))$ where $\mu^+(p) \in \mathbb{N} \cup \{0\},\mu^-(p) \in \mathbb{N} \backslash 2\mathbb{N} $.
			\end{enumerate}
			Hence, for all $p_i$, $\mu^+(p_i) \ \in \ \mathbb{N} \cup \{0\} \ \& \\ $
			$\mu^-(p_i) \ \in \ 
			\begin{cases}
			2\mathbb{N} \cup \{0\} &, \  \textnormal{if} \ p_i \ \textnormal{has positive incident arcs}. \\
			\mathbb{N} \backslash 2\mathbb{N} &, \  \textnormal{if} \ p_i \ \textnormal{has negative incident arcs}. \\
			\end{cases}
			$\\
		\end{enumerate}
		\item \textbf{Transitions are negatively signed.}\\
		By following the same procedure as in the case when transitions are positively signed, we find that for all $p_i$, $\mu^+(p_i) \ \in \ \mathbb{N} \cup \{0\} \ \& \\ $
		$\mu^-(p_i) \ \in \ 
		\begin{cases}
		
		\mathbb{N} \backslash 2\mathbb{N} &, \  \textnormal{if} \ p_i \ \textnormal{has positive incident arcs}. \\
		2\mathbb{N} \cup \{0\} &, \  \textnormal{if} \ p_i \ \textnormal{has negative incident arcs}. \\
	\end{cases}
	$\\

\end{enumerate}
Thus, we get a marking $\mu$ w.r.t. which $P$ and $T$ are dominating sets.

\end{proof}

\begin{remark}
The above theorem may be used in case of a disaster to check which transitions representing events in the system will dominate and hence may be used for preparedness against any disaster.
\end{remark}
\begin{remark}
Theorems 1 and 2 given above show that:
\begin{enumerate}
	\item We can get a dominating set if we begin with the specified initial marking as mentioned in the proof.
	\item $P \ \& \ T$ sets in a PN represent conditions and events respectively of the system modeled.So,we can check whether conditions or events dominate in the given system w.r.t. a given marking.
	\item If we know the structure of an SPN and marking $\mu_0$, then we can check whether domination can occur by checking if $\exists \  \mu \ \in \ R(N^*,\mu_0)$ w.r.t. which there exists a dominating set.So, in order to avoid or force domination we can restrict (or force) SPN to avoid (or reach) such a marking.
\end{enumerate}
\end{remark}
\section{Applications of Domination}
We discuss applications of domination in SPN that can be utilized in various areas.
\subsection{Producer -Consumer Problem}
\begin{figure}[ht]
\centering
\includegraphics[scale=0.5]{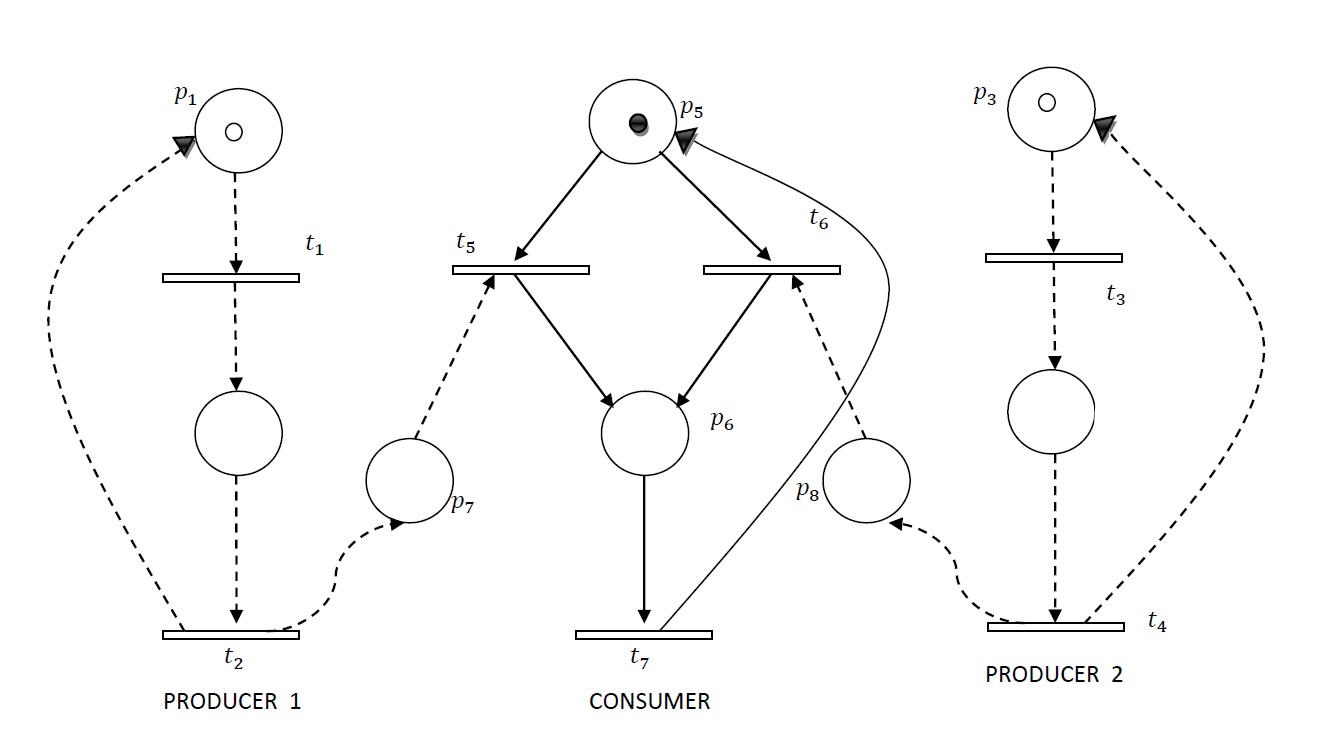}
\caption{\textit{Producer-Consumer Problem}}
\label{fig9}
\end{figure}

Consider, a standard Producer -Consumer Problem with two producers producing a same product (assuming quality,price and other conditions are same),we need to check whether one of the producer can dominate the market over the other.This can happen due to availability of product is greater for one producer as compared to other or because one product is well known due to its better marketing, etc.
Consider an SPN model for the problem given in figure \ref{fig9}.

Now,in order to check if producer 2 dominates the market ,we need to find a set of vertices $D_1$ which is a dependent dominating set over a maximal set of markings $M$. A dependent dominating set is considered because one producer is said to dominate over the other if such a domination exists over a period of time, not for just an instant.

Choose $D_1=V \backslash \{p_7,t_7\}$.Find if there exists a set of markings $M$(maximal) with marking $\mu \in M$ such that $\mu(p_6)\neq 0$ and $D_1$ must be a dominating set w.r.t. $M$. Similarly, to check if producer 1 dominates the market we need to check domination of set $D_2$ w.r.t. a set of markings where $D_2 =V \backslash \{p_8,t_7\}$.
\subsection{Search of food by bees}
\begin{figure}[ht]
\centering
\includegraphics[scale=0.32]{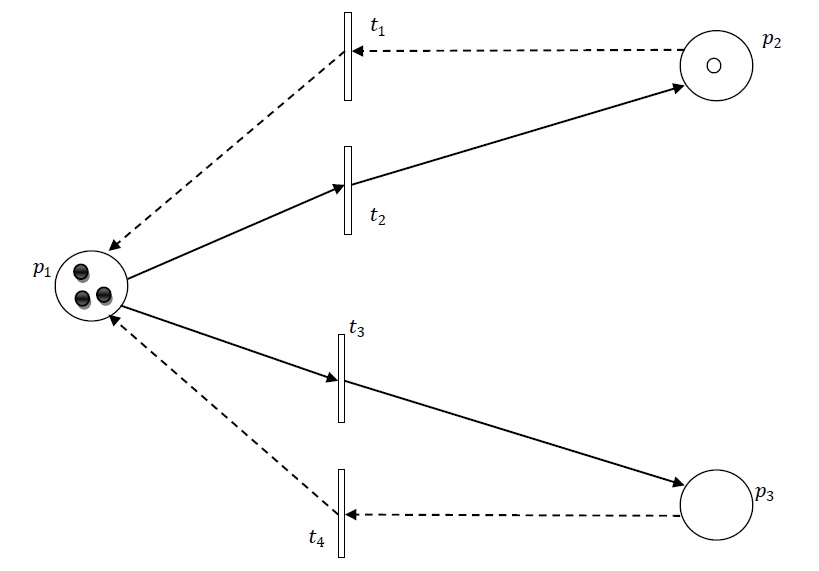}
\caption{\textit{An SPN model with initial marking $((3,0,0),(0,1,0))$ for food finding by Scout Bees}}
\label{fig2}
\end{figure}

Bees (Scout bees) go out in search of food.The one which find the food will return to the hive and celebrate.This scout bee can be considered to dominate other scout bees.This problem can be modeled using an SPN and then it can be identified which scout bee will dominate the bee-hive.Consider the figure \ref{fig2} where place $p_1$ represents the bee-hive while $p_2,p_3$ represent possible food locations where scout bees $A$ and $B$ respectively search for food.The transitions $t_2,t_3$ represent events of food search while $t_1,t_4$ represent events of food search completion.The positive tokens are used to represent bees and negative ones to represent food.
\par According to the initial marking of SPN, location $p_2$ has food while location $p_3$ does not.Therefore, bee $A$ must dominate.This can be verified by checking that the set $D_1=\{p_1,p_3,t_1,t_2,t_3,t_4\}$ is a dominating set w.r.t.the initial marking $\mu_0$ rather than set $D_2=\{p_1,p_2,t_1,t_2,t_3,t_4\}$.In the later case, bee $B$ will dominate the bee-hive.
The model can be extended if more than two bees search for food.
\subsection{Finding papers with similarity to a given paper using softwares like Turnitin}
\begin{figure}[ht]
\centering
\includegraphics[scale=0.3]{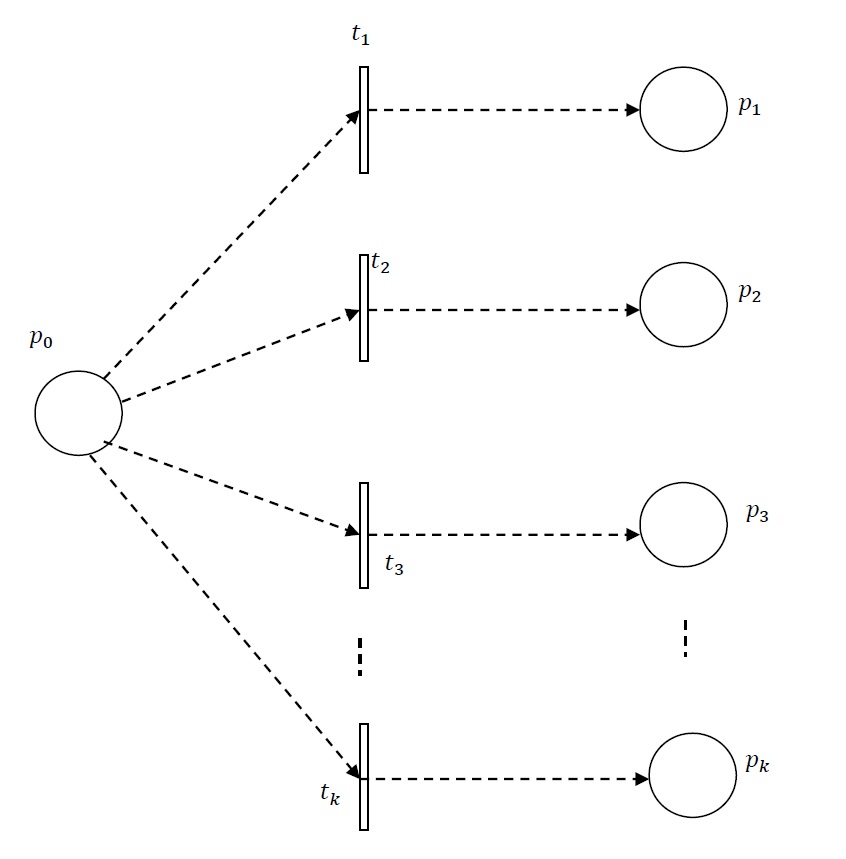}
\caption{\textit{Finding similarity in Papers}}
\label{fig3}
\end{figure}

Consider a paper which need to be checked for similarity using a software.This paper is compared with web content that is publicly available,  books, papers in journals, articles and other content which is present in a software database.Let the paper to be checked be represented by a place $p_0$ and the rest of the content be represented by places $p_1,p_2,p_3,\ldots p_k$.To check for similarity, an SPN model is formed by connecting place $p_0$ to all other places via transitions $t_1,t_2,t_3,\ldots t_k$ and by using negative arcs as in figure \ref{fig3}.\par
While comparing paper $p_0$ with another article represented by place $p_i$ (say),$1\leq i \leq k$, a matching algorithm is used to find a set of strings within submitted paper $p_0$ that matches with the papers maintained in its database.If a similarity exists, a negative token is generated in place $p_0$ which can be used to fire corresponding transition $t_i$.In this way, the submitted paper is checked for similarity with all the content present in the database.After the comparisons are completed, we get a new marking for the SPN in which all the articles that have some similarity with the submitted paper get a negative token in the place representing it.All such places will form a list of articles that are similar to the paper submitted which is to be tested for similarity.
\par Now, in order to find the list of all articles which have some similarity with the submitted paper,the concept of domination can be used instead of finding all the places having a negative token.
Begin with set $D_1=P=\{p_0,p_1,p_2,\ldots p_k\}$.Check whether set $D_1$ is a dominating set w.r.t. the final marking(say $\mu'$) obtained after the matching algorithm is complete.If yes,then all the papers $p_i,1\leq i\leq k$ have some similarity with the submitted paper.If not, we find $D_2 \subseteq T$ such that $D'=D_1 \cup D_2$ is a dominating set w.r.t.$\mu'$.Then ,the set $\{p_i|t_i \in T\backslash D_2\}$ will form the set of all the articles which have similarity with the submitted paper.

\bibliographystyle{abbrvnat}
\bibliography{mybibfile}
\end{document}